\documentclass[12pt]{article}

\usepackage[utf8]{inputenc}

\usepackage{amsmath}
\usepackage{amsfonts}
\usepackage{cases}
\usepackage[colorlinks]{hyperref}
\hypersetup{
  colorlinks=true,
  linkcolor=red,
  citecolor=red
}
\AtBeginDocument{%
  \hypersetup{%
    linkcolor=blue,%
    citecolor=blue,%
  }%
}%

\begin{document}

%--------------------------------------------------------------------------

\begin{center}
{\Large
	{\sc Extension de la régression linéaire généralisée sur composantes supervisées (SCGLR) aux données groupées.}
}
\bigskip

 Jocelyn Chauvet $^{1}$ \& Catherine Trottier $^{1}$ \& Xavier Bry $^{1}$ \& Frédéric Mortier $^{2}$  
\bigskip

{\it
$^{1}$ Institut Montpelli\'erain Alexander Grothendieck, CNRS, Univ. Montpellier, France. 
jocelyn.chauvet@umontpellier.fr, catherine.trottier@univ-montp3.fr, xavier.bry@univ-montp2.fr

$^{2}$ Cirad, UR BSEF, Campus International de Baillarguet - TA C-105/D - 34398 Montpellier, frederic.mortier@cirad.fr
}
\end{center}
\bigskip

%--------------------------------------------------------------------------

{\bf R\'esum\'e.} 
Nous proposons de construire des composantes permettant de régulariser un Modèle Linéaire Généralisé Mixte (GL2M) multivarié. Un ensemble de réponses aléatoires $Y$ est modélisé par un GL2M, au moyen d'un ensemble $X$ de variables explicatives, et d'un ensemble $T$ de variables additionnelles. Les variables explicatives dans $X$ sont supposées nombreuses et redondantes : il est donc nécessaire de régulariser la régression linéaire généralisée mixte. À l'inverse, les variables de $T$ sont supposées peu nombreuses et sélectionnées de sorte à n'exiger aucune régularisation. 
La régularisation consiste ici à construire un nombre approprié de composantes orthogonales permettant tout à la fois une bonne modélisation de $Y$ et l'extraction d'informations structurelles dans $X$. Pour cela, nous proposons d'insérer à chaque étape de l'algorithme de Schall permettant l'estimation d'un GL2M, l'optimisation d'un critère propre à SCGLR.
Cette extension de la méthode SCGLR est testée et comparée à d'autres méthodes de régularisation de type Ridge et Lasso, sur données simulées et réelles. 

\smallskip

{\bf Mots-cl\'es.} Modèles à composantes, GL2M multivarié, Effet aléatoire, Pertinence structurelle, Régularisation, SCGLR.
\bigskip\bigskip

{\bf Abstract.}
We address component-based regularisation of a multivariate Generalized Linear Mixed Model.  
A set of random responses $Y$ is modelled by a GLMM, using a set $X$ of explanatory variables and a set $T$ of additional covariates.
Variables in $X$ are assumed many and redundant : generalized linear mixed regression demands regularisation with respect to $X$. 
By contrast, variables in $T$ are assumed few and selected so as to demand no regularisation.
Regularisation is performed building an appropriate number of orthogonal components that both contribute to model $Y$ and capture relevant structural information in $X$.
We propose to optimize a SCGLR-specific criterion within a Schall's algorithm in order to estimate the model.
This extension of SCGLR is tested on simulated and real data, and compared to  Ridge- and Lasso-based regularisations.

\smallskip


%--------------------------------------------------------------------------

\section{Données, modélisation et problème.}

\paragraph{}
Nous considérons un ensemble de $q$ réponses aléatoires 
$Y_{n \times q} = \left\lbrace y^1, \ldots, y^q \right\rbrace$, 
expliquées par deux ensembles de covariables 
$X_{n \times p} = \left\lbrace x^1, \ldots, x^p \right\rbrace$ et 
$T_{n \times r} = \left\lbrace t^1, \ldots, t^r \right\rbrace$ aux statuts bien différents.
Les $p$ variables explicatives contenues dans $X$ sont supposées nombreuses et redondantes tandis que les $r$ variables additionnelles contenues dans $T$ sont supposées peu nombreuses et sélectionnées de sorte à éviter les redondances. 
Les variables dans $T$ peuvent alors être conservées dans le modèle sans traitement particulier. Par contre, on suppose que seulement quelques dimensions $H<p$ dans $X$ suffisent à capter la majorité de l'information utile pour modéliser $Y$.
En se basant sur les structures fortes de $X$, il s'agira de construire un nombre approprié de composantes orthogonales, dans le but de prédire au mieux $Y$. 

\paragraph{}
Considérant des distributions de réponses aléatoires appartenant à la famille exponentielle, chacune des $y^k$ est modélisée selon un GL2M (\hyperlink{McCulloch}{McCulloch, Searle (2001)}). 
Plus précisément, les situations considérées sont celles où les $n$ unités statistiques ne sont pas indépendantes mais structurées en $N$ groupes. L'effet aléatoire introduit aura précisément pour objectif de modéliser la dépendance à l'intérieur de chacun des groupes.
Précisons de plus que les réponses $Y$ sont supposées indépendantes conditionnellement à $X \cup T$.

\paragraph{}
De nombreuses méthodes de régularisation dans le cadre de la modélisation GLM 
(\hyperlink{McCullagh}{McCullagh, Nelder (1989)}) 
ont été développées ces dernières années. 
Dans le cadre univarié, c'est à dire lorsque $Y = \left\lbrace y \right\rbrace$, 
\hyperlink{Bastien}{Bastien et al. (2004)} 
proposent de combiner, à la manière de PLS, les régressions linéaires généralisées de la variable dépendante sur chacun des régresseurs pris isolément. Cependant, la technique en question, appelée PLSGLR, semble insatisfaisante car elle ne tient pas compte de la variance induite par la modélisation GLM. 
Toujours dans le cadre univarié, 
\hyperlink{Marx}{Marx (1996)} 
introduit la stratégie nommée IRPLS, qui tient compte de la matrice de poids des observations provenant de la modélisation GLM pour construire les composantes PLS. 
Dans la lignée de \hyperlink{Marx}{Marx (1996)}, 
\hyperlink{Bry}{Bry, Trottier et al. (2013)} proposent SCGLR, pour Régression Linéaire Généralisée sur Composantes Supervisées. Cette extension multivariée de la méthode précédente recherche des composantes communes à tous les $y^k$. Pour les construire, un nouveau critère est introduit et maximisé dans chacune des étapes de l'algorithme des scores de Fisher (FSA). 

\paragraph{}
Cependant, toutes ces méthodes se fondent sur une hypothèse importante : l'indépen-dance des $n$ unités statistiques intervenant dans la modélisation. On se propose ici d'étendre la méthode SCGLR en y introduisant une structure de dépendance entre les unités statistiques via un effet aléatoire. 

\paragraph{}
L'une des motivations de ce travail est la modélisation et la prédiction de l'abondance d'espèces d'arbres dans la forêt tropicale du bassin du Congo. En effet, les mesures d'abondance d'espèces étant naturellement organisées en groupes dans l'espace (les concessions forestières dans nos données), la structure de dépendance induite doit être prise en compte dans les modèles.

\section{Retour sur SCGLR avec variables additionnelles.}

\paragraph{}
Nous nous plaçons dans cette partie dans le cas où chacune des variables réponses $y^k$ est modélisée selon un GLM. Pour simplifier, nous ne présenterons que la recherche de la première composante ($H=1$).

\paragraph{}
Le premier fondement conceptuel de SCGLR consiste à chercher dans $X$ une composante commune  notée $f=Xu$, optimale pour l'ensemble des $y^k$. Les prédicteurs du FSA sont donc partiellement colinéaires : 
\begin{center}
$\forall k \in \left\lbrace 1, \ldots, q\right\rbrace, 
\eta^k = (Xu)\gamma_k + T\delta_k$, 
\end{center}
avec, pour des questions d'identification, $u'Au=1$ où $A$ est une matrice symétrique définie positive. 
Si l'on suppose l'indépendance des réponses $y^k$ et des $n$ unités statistiques, la vraisemblance s'écrit simplement :
\begin{equation*}
L(y|\eta) = \prod_{i=1}^n \prod_{k=1}^q L_k (y_i^k | \eta_i^k).
\end{equation*}
En raison du terme $\gamma_k u$, le modèle linéarisé construit à chaque étape du FSA n'est pas linéaire : une procédure de moindres carrés alternés est donc adoptée. 
En notant $z^k$ les variables de travail classiques du FSA et $W_k^{-1}$ leurs matrices de variance, le programme suivant doit être considéré :
\begin{equation*}
\mathcal{Q}_1 : \underset{u'Au=1}{max} \; \psi_T(u), \quad
\text{avec} \quad \psi_T(u) = \sum_{k=1}^q \Vert z^k \Vert_{W_k}^2 \cos_{W_k}^2 
(z^k, \langle Xu, T\rangle).
\end{equation*}

\paragraph{}
Le deuxième fondement conceptuel de SCGLR consiste à introduire une mesure de la proximité de la composante $f=Xu$ aux structures fortes de $X$ : la pertinence structurelle. 
Avec $W$ la matrice des poids des $n$ unités statistiques, la composante la plus pertinente serait solution du programme :
\begin{equation*}
\mathcal{Q}_2 : \underset{u'Au=1}{max} \; \phi(u), \quad
\text{avec} \quad
\phi(u) 
= \left(    {\overset{p}{\underset{j=1}{\sum}}}  \langle Xu | x^j \rangle_{W}^{2l}      
\right)^{\frac{1}{l}}
= \left(   {\overset{p}{\underset{j=1}{\sum}}} 
(u' X' W x^j {x^j}' W X u)^{l}      
\right)^{\frac{1}{l}}, 
\end{equation*}
le paramètre $l \in \left[ 1, +\infty \right] $ permettant de régler la localité des faisceaux de variables visés.

\paragraph{}
La stratégie SCGLR se propose alors de faire une synthèse entre les programmes $\mathcal{Q}_1$ et $\mathcal{Q}_2$, et de choisir la composante $f=Xu$ solution du programme :
\begin{equation*}
\mathcal{Q} : \underset{u'Au=1}{max} \; \left[\phi(u)\right]^s \: \left[\psi_T(u)\right]^{1-s}. 
\end{equation*}
Le paramètre $s$ permet de régler l'importance relative accordée à la pertinence structurelle (programme $\mathcal{Q}_2$) par rapport à la qualité d'ajustement (programme $\mathcal{Q}_1$).

\section{Adaptation de SCGLR aux données groupées.}

\paragraph{}
Nous proposons une adaptation de la méthode SCGLR aux données groupées, pour lesquelles l'hypothèse d'indépendance des unités statistiques n'est plus valable. On peut penser par exemple aux données structurées en groupes dans l'espace ou aux données répétées dans le temps (données longitudinales). Nous modélisons la dépendance induite par un effet aléatoire, d'où une modélisation via un GL2M de chacune des réponses $y^k$.

\subsection{Construction de la première composante.}

\paragraph{}
Nous proposons de conserver la propriété de colinéarité partielle (sur $X$) des prédicteurs. L'injection d'un effet aléatoire dans chaque prédicteur permettra de traduire la structure de dépendance des unités statistiques. La matrice $U$ contenant les données d'appartenance aux groupes, les prédicteurs s'écrivent alors :
\begin{center}
$\forall k \in \left\lbrace 1, \ldots, q\right\rbrace, \quad
\eta_{\xi}^k = (Xu)\gamma_k + T\delta_k + U\xi_k \;$, 
avec $\; u'Au=1$.
\end{center}
Les effets aléatoires $\xi_1, \ldots, \xi_q$ sont supposés indépendants, et pour tout $\: k \in \left\lbrace 1, \ldots, q\right\rbrace$, $\xi_k$ est supposé distribué selon une loi $\mathcal{N}_N(0, D_k = \sigma_k^2 Id)$, avec $N$ le nombre de groupes.

\paragraph{}
La complexification de la structure de variance ne permet plus d'appliquer l'algorithme des scores de Fisher, qui suppose l'indépendance des unités statistiques.
Nous avons alors envisagé d'adapter l'algorithme de 
\hyperlink{Schall}{Schall (1991)} 
en y introduisant, à chaque étape, la procédure alternée suivante  : 
\begin{itemize}
\item
À $u$ fixé, on estime les paramètres $\gamma_k$, $\delta_k$ et $\sigma_{k}^{2}$ par un système de Henderson 
(\hyperlink{Henderson}{Henderson (1975)}).
\item
À $\gamma_k$, $\delta_k$ et $\sigma_{k}^2$ fixés, on construit la composante $f=Xu$.
\end{itemize}

\paragraph{}
En effet, les variables de travail qui interviennent dans notre algorithme peuvent s'écrire sous la forme :
\begin{center}
$z_{\xi}^k = (Xu)\gamma_k + T\delta_k + U\xi_k + e^k$. 
\end{center}
À $u$ fixé, les variables de travail $z_{\xi}^k$ sont considérées, à la manière de Schall, comme modélisées par un L2M. Une estimation courante des paramètres $\gamma_k$, $\delta_k$ et $\sigma_{k}^2$ est alors accessible par la résolution d'un système de Henderson, dont une justification dans ce cadre se trouve notamment dans 
\hyperlink{Stiratelli}{Stiratelli et al. (1984)}, 
au moyen d'un raisonnement bayésien. 

\paragraph{}
Quant à la construction de la composante $f=Xu$, elle est toujours solution d'un programme pouvant se mettre sous la forme $\mathcal{Q}$, mais il est nécessaire de modifier l'expression $\psi_T(u)$. Conditionnellement à l'effet aléatoire $\xi_k$, la variable de travail $z_{\xi}^k$ possède une structure de GLM. Nous proposons alors de modifier le critère de qualité d'ajustement en tenant compte de la variance de $z_{\xi}^k$ conditionnellement à $\xi_k$. Pour cela, nous posons :
\begin{equation*}
\psi_T(u) = \sum_{k=1}^q \Vert z_{\xi}^k \Vert_{W_{\xi,k}}^2 \cos_{W_{\xi,k}}^2 (z_{\xi}^k, \langle Xu, T \rangle), 
\quad \text{avec} \quad W_{\xi,k}^{-1} := \mathbb{V}(z_{\xi}^k | \xi_k).
\end{equation*}

\subsection{Construction des composantes suivantes.}

\paragraph{}
Supposons avoir construit les $h$ premières composantes, et posons 
$F^h = \left\lbrace f^1, \ldots, f^h \right\rbrace$. 
La $(h+1)^{\text{ème}}$ composante est conçue pour compléter au mieux les $h$ composantes précéden-tes et $T$, autrement dit $T^h = F^h \cup T$. De plus, nous imposons l'orthogonalité entre $f^{h+1}$ et $F^h$, via la contrainte :
\begin{center}
${F^h}' W f^{h+1} = 0.$
\end{center}
La composante $f^{h+1} = Xu^{h+1}$ est donc obtenue en résolvant le programme suivant, dont $u^{h+1}$ est solution :
\begin{equation*}
\begin{cases}
\textbf{Maximiser} \quad  \left[\phi(u)\right]^{s} \: \left[\psi_{T^h}(u)\right]^{1-s}    \\
\textbf{sous les contraintes : }  \quad u' A u = 1 \; \text{et} \; ({F^h}' WX) u = 0
\end{cases}
\end{equation*}

\section{Tests numériques.}

\paragraph{}
Pour illustrer le potentiel de la méthode, nous présenterons dans un premier temps des résultats sur données simulées et réelles impliquant une modélisation GL2M (notamment des données groupées poissonniennes, lien log).

Des comparaisons seront également effectuées avec d'autres méthodes de régularisation qui, elles, ne font pas appel à la construction de variables latentes (nos composantes), mais consistent à appliquer une pénalisation sur la norme des coefficients de régression. Deux de ces méthodes ont été retenues :
\begin{itemize}
\item
Celle proposée par 
\hyperlink{Eliot}{Eliot et al. (2011)}, 
qui étend la régression ridge au cas des modèles linéaires mixtes (L2M) univariés, non généralisés.
\item
Celle proposée par 
\hyperlink{Groll}{Groll et Tutz (2014)}, 
qui consiste à injecter une pénalisation en norme $L^1$ lors de l'ajustement de modèles GL2M, toujours dans le cadre univarié.
\end{itemize}

\section{Conclusion.}

\paragraph{}
L'extension de SCGLR proposée est un bon compromis entre une modélisation GL2M (possédant de bonnes qualités prédictives dans le cadre de données groupées mais très instable si les régresseurs sont fortement redondants) et des stratégies de type GLM sur composantes principales (qui ne tiennent pas compte des variables réponses pour la construction des composantes).
Les qualités de SCGLR sont préservées, et l'extension présentée permet de réagir face à des données groupées en produisant des modèles prédictifs robustes basés sur des composantes interprétables.

%Quelques rappels :
%%
%\begin{center}
%%
%\begin{tabular}{lr} \hline
%%
%Accent aigu :              &  \'e; \\
%Accent grave :             &  \`a;\\
%Accent circonflexe :       &  \^o mais \^{\i};\\
%Tr\'emas :                 &  \"o mais \"{\i};\\
%C\'edille :                &  \c{c}. \\ \hline
%\end{tabular}
%%
%\end{center}

%--------------------------------------------------------------------------

\section*{Bibliographie}
\label{Bibliographie}
\hypertarget{McCulloch}{
\noindent [1] C.E. McCulloch, S.R. Searle (2001). \textit{Generalized, Linear, and Mixed Models}, John Wiley \& Sons, New-York, USA.}

\medskip

\hypertarget{McCullagh}{
\noindent [2] P. McCullagh, J.A. Nelder (1989). \textit{Generalized linear models}, Chapman and Hall, New York, USA.}

\medskip

\hypertarget{Bastien}{
\noindent [3] P. Bastien, V. Esposito Vinzi, M. Tenenhaus (2004). \textit{PLS generalized linear regression}, Computational Statistics \& Data Analysis, 48(1) :17-46.}

\medskip

\hypertarget{Marx}{
\noindent [4] B. D. Marx (1996). \textit{Iteratively reweighted partial least squares estimation for generalized linear regression}, Technometrics, 38(4) :374–381.}

\medskip

\hypertarget{Bry}{
\noindent [5] X. Bry, C. Trottier, T. Verron, F. Mortier (2013). \textit{Supervised component generalized linear regression using a PLS-extension of the Fisher scoring algorithm}, Journal of Multivariate Analysis, 119(C) :47–60.}

\medskip

\hypertarget{Schall}{
\noindent [6] R. Schall (1991). \textit{Estimation in generalized linear models with random effects}, Biome-trika, 78(4) :719–727.}

\medskip

\hypertarget{Henderson}{
\noindent [7] C.R. Henderson (1975). \textit{Best linear unbiaised estimators and prediction under a selection model}, Biometrics, 31 :423–447.}

\medskip

\hypertarget{Stiratelli}{
\noindent [8] R. Stiratelli, N. Laird, J.H. Ware. \textit{Random-Effects Models for Serial Observations with Binary Response}, Biometrics, 40(4) :961-971.}

\medskip

\hypertarget{Eliot}{
\noindent [9] M. Eliot, J. Ferguson, M.P. Reilly and A.S. Foulkes (2011). \textit{Ridge Regression for Longitudinal Biomarker Data}, The International Journal of Biostatistics, 7(1) :1-11.}

\medskip

\hypertarget{Groll}{
\noindent [10] A. Groll, G. Tutz (2014). \textit{Variable Selection for Generalized Linear Mixed Models by L1-Penalized Estimation},  Statistics and Computing, 24(2) :137-154.}
\end{document}